# Demand, Supply, and Performance of Street-Hail Taxi

Ruda Zhang and Roger Ghanem

*Abstract*—Travel decisions are fundamental to understanding human mobility, urban economy, and sustainability, but measuring it is challenging and controversial. Previous studies of taxis are limited to taxi stands or hail markets at aggregate spatial units. Here we estimate the dynamic demand and supply of taxis in New York City (NYC) at street segment level, using in-vehicle Global Positioning System (GPS) data which preserve individual privacy. To this end, we model taxi demand and supply as non-stationary Poisson random fields on the road network, and pickups result from income-maximizing drivers searching for impatient passengers. With 868 million trip records of all 13,237 licensed taxis in NYC in 2009 – 2013, we show that while taxi demand are almost the same in 2011 and 2012, it declined about 2% in spring 2013, possibly caused by transportation network companies (TNCs) and fare raise. Contrary to common impression, street-hail taxis out-perform TNCs such as Uber in high-demand locations, suggesting a taxi/TNC regulation change to reduce congestion and pollution. We show that our demand estimates are stable at different supply levels and across years, a property not observed in existing matching functions. We also validate that taxi pickups can be modeled as Poisson processes. Our method is thus simple, feasible, and reliable in estimating street-hail taxi activities at a high spatial resolution; it helps quantify the ongoing discussion on congestion charges to taxis and TNCs.

*Index Terms*—Urban computing, spatial network, matching function, search friction, congestion.

## I. Introduction

TAXI transportation constitutes a key component of urban mobility, and is a rare case of industrial economy where detailed and comprehensive transaction records are available. As metropolitan areas over the world continue to grow in population and economy, traffic congestion and induced pollution have got worse. Average traffic delay in Manhattan Central Business District increased 28% between 2013 and 2017 [1], where an important source is growth in taxi or transportation network company (TNC) vehicles, especially unoccupied ones [2]. Increasing search efficiency is thus key to alleviate the ongoing increase in congestion. To understand taxi operation, one needs to quantify the spatial-temporal patterns of taxi activities. More critically, while pickup and drop-off information are easily measured [3], taxi supply and demand levels are often of greater interest [4], [5]. Unfortunately, such attributes are hard to obtain: taxi supply estimation requires high sampling rate GPS trajectories [6], which is challenging to store, transmit, and process; taxi demand is impossible to measure without pervasive instrumentation on the entire population because all are potential taxi passengers. Estimating these unmeasured attributes can complement our insight into urban mobility obtained from other sources such as smart-card data [7] and mobile phone data [8], [9]. It also enables studies of supply-demand relation, and competition between taxis and TNCs [10]. Governments can adjust taxi and TNC regulations to improve transportation efficiency, and transportation operators can benefit from this information to improve service quality. This paper is thus motivated to study the supply and demand distributions of street-hail taxi, and their implication on performance.

In this paper, supply refers to a vacant taxi in service searching for passengers, and demand refers to a potential passenger, or hailer, who tries to hail a taxi on the street, regardless the eventual mode of transportation. We note that the supply and demand defined here are not those of a homogeneous good in economics vocabulary, but actually two factors of production. We choose these terms to conform with common understanding. This distinction is important because previous studies of taxi service have assumed market clearing at economic equilibrium, which implies all hailers are eventually picked up [11], [12].

In urban computing and trajectory mining literature, studies using taxi GPS records largely started around 2010. Vacant taxi availability has been studied by [13], [14]. Many predictive models of taxi demand have been proposed, despite treating actual pickups as demand although the latter is obviously larger than the former, see for example [15]–[18]. Taxi demand and pickups were distinguished in [19], the first among similar works. Few studies used road network to estimate road capacity with a speed threshold [20], estimate traffic speed [21], [22], or route shared taxi rides [3], [23]. Many papers have proposed recommender systems to improve taxi driver search efficiency [14], [24]–[28], which invariantly use observed pickups in calculating pickup probability. While this approach is convenient, it ignores the interaction between demand and supply in the resulting pickups, thus undermining model effectiveness. Understanding such interaction is a main purpose of this paper, and our models complement the aforementioned studies.

In taxi search friction and matching literature, matching functions are commonly used to relate demand and supply

Manuscript received July 5, 2018; revised December 31, 2018 and June 27, 2019; accepted August 28, 2019. This work was supported by the National Science Foundation, Resilient Interdependent Infrastructure Processes and Systems, under Grant 14-524. The Associate Editor for this article was S. Sacone. *(Corresponding author: Ruda Zhang.)*

The authors are with the Department of Civil and Environmental Engineering, University of Southern California, Los Angeles, CA 90089 USA (e-mail: rudazhan@usc.edu; ghanem@usc.edu).

This article has supplementary downloadable material available at http://ieeexplore.ieee.org, provided by the authors.

Digital Object Identifier 10.1109/TITS.2019.2938762

with matches. But the functions used are either aggregate models or obtained by simulation, and not capturing the characteristics of street-hail taxis. For example, [29] used the minimum number of hailers and vacant taxis as the number of matches, assuming all buyers and sellers at the same place match simultaneously. Reference [12], [30]–[32] used the Cobb-Douglas production function, where meeting rate is a power function of the numbers of hailers and vacant taxis; [32] also captures traffic speed and discussed the effect of heterogeneous traffic congestion on taxi dynamics. Reference [33] used a matching function obtained from numerical simulation to estimate the total number of hailers in each hour of a typical weekday. Their matching function maps from the numbers of hailers and vacant taxis to taxi search time, and they estimated demand by interpolating and inverting the function. However, in their simulation, hailers arrive in batches at fixed time intervals, distributed with uniform probability on grid nodes of each of eight areas in core Manhattan, and wait indefinitely while vacant taxis move as random walkers. Reference [34] estimates supply and demand distributions over 39 clusters of census tracts for every 5-minute periods between 6am and 4pm weekdays. Equilibrium supply distribution is solved for each period based on previous values, the movement of occupied taxis, and driver decision to maximize net revenue. He uses the equilibrium matching function derived in [35], which is based on [36]'s urn-ball matching problem, that is, taxis have capacity one and customers arriving at a matched taxi are rejected.

We build probabilistic models of taxi pickup on street segments, which are related to queueing models to get analytical forms of the matching functions (Subsection II-A). Queueing models have long been applied to pickup at taxi stands [37], where either taxis or passengers can be waiting, if there is any. We show that queueing theory [38]–[41] can also be adapted to hail markets where taxis search on the street network. The choice of street segment as spatial unit, compared with areas and points, respects the data generating process, and allows for robust spatial aggregation and interpretable mechanistic models. To our knowledge, this is the first analytical model of taxi–passenger matching on street segments, which applies to any urban area with street-hail taxis. The matching function associated with our model is thus mechanistic, and is shown to provide reliable demand estimates while existing models cannot. Like many taxi-related research, our model assumes Poisson arrivals, but our paper appears to be the first to validate this assumption. The interaction among taxi supply, demand, and pickups captured in our model complements previous works that conflate pickup with demand.

Since taxi has long been a government regulated segment of urban transportation, for cities that have been collecting taxi GPS trajectory data, our model can be readily applied to estimate taxi demand. If for technical or historical reasons only trip origin-destination locations are available, we also provide a simple formula to estimate the equilibrium supply route distribution (Subsection II-B). To showcase our method, we estimated the spatial-temporal distributions of taxi supply and demand in Manhattan, NYC (Section III). As ride-hiring markets in cities worldwide have been fragmented by TNCs, it is harder now to estimate the complete demand distribution from a single source of data. But our method still applies to the demand seeking street-side service. And by comparing estimated taxi demand distributions before and after TNC entry, we can have a detailed understanding of TNCs' impact on taxi transportation. We show that street-hail taxis perform better than TNCs in high demand locations, which can inform current taxi and TNC regulation, such as the congestion charge proposal in NYC.

## II. Street-Hail Taxi Operation

We regard transportation activities as periodic non-stationary random fields. In other words, events differ at different places, change over time, and would not be the same if there were independent duplicates of the world, but the same random entity occurs at regular occasions in time. Temporal regularity is the key to understand and estimate such random entities. In the case of taxi transportation, system states can be modeled at equilibrium when environmental conditions such as traffic speed, passenger demand, and driver supply are held stationary by observing at regular, short time windows.

### A. Segment-Level Pickup Models

To study the kinetics of street-hail taxi, we dissect a street network at intersections into street segments, and propose a class of segment-level pickup models, specified by $(A, B, C, D)$:

1) Hailers arrive at a one-directional street segment $x$ of length $l$ as a stochastic process in time-space $\mathbb{R}_+ \times [0, l]$, with independent inter-arrival time distribution $A$ whose occurrence rate is called demand rate $\mu_d$, and uniform spatial distribution along the segment. A group of hailers traveling together are counted as one.
2) Vacant taxis enter the segment as a stochastic process in time $\mathbb{R}_+$, with independent inter-arrival time distribution $B$ whose occurrence rate is called supply rate $\mu_s$.
3) Hailer patience, or maximum waiting time, $T$ is distributed as $C$. We call $\mathbb{E}T$ hailer mean patience, and define impatience $\mu_t = 1/\mathbb{E}T$.
4) In case multiple hailers are present when a vacant taxi arrives, either of the following pickup disciplines $D$ may be used: greedy ($G$), the driver picks up the hailer closest to segment entrance; courteous ($C$), the hailer who has waited for the longest time gets in the taxi.

Fig. 1A shows a diagram of the pickup model. We assume that taxi drivers do not deny hailers, so if a vacant taxi enters the segment while at least one hailer is waiting, the driver will pick up a hailer. Note that although passenger denial exists in reality, the proportion of denied passengers is not significant [42]. This assumption is only needed to preserve homogeneity among drivers, so its impact is further weakened if taxi drivers have similar denial patterns.

This model establishes taxi pickup as another stochastic process in time and space, but we are mainly interested in the pickup rate $\mu_p$. Given model specification $(A, B, C, D)$, $\mu_p$ is a function of the model parameters: $\mu_p(\mu_d, \mu_s, \mu_t)$. Due to the scaling property of time, the pickup rate function is a homogeneous function of degree one: $\mu_p(\alpha\mu_d, \alpha\mu_s, \alpha\mu_t) = \alpha\mu_p(\mu_d, \mu_s, \mu_t)$. So we reduce the pickup rate function to $\sigma(\rho, \kappa)$ using three independent dimensionless quantities of the model: demand fulfillment $\sigma = \mu_p/\mu_d$, supply-demand ratio $\rho = \mu_s/\mu_d$, and cover number $\kappa = \mu_t/\mu_d$. Demand fulfillment is the proportion of taxi demand fulfilled; supply-demand ratio is the number of vacant taxis passed before a new hailer is expected to appear; cover number can be interpreted as the multiples of hailer mean patience to cover the expected hailer inter-arrival time. Relating to queueing models, several specifications have pickup rate functions in analytical form, see Eq. 1–7 and Fig. 1B. For a taxi system, both hailer waiting time and taxi vacancy rate should be low to reduce cost of time, and unnecessary congestion and pollution. To quantify the performance of street-hail taxi, we interpret deterministic hailer patience $T$ as service guarantee, interpret demand fulfillment $\sigma$ as quality of service, and define supply realization $\varsigma = \mu_p/\mu_s$ as the proportion of supply matched, then the pickup rate function $\mu_p(\mu_d, \mu_s, \mu_t)$ can be rewritten as the relationship among these performance measures and demand rate $\mu_d$: $\sigma(\varsigma, T, \mu_d)$, see Fig. 1C. The trade-off between demand fulfillment $\sigma$ and supply realization $\varsigma$ depends on the city planner, but Pareto efficiency always improves as demand rate $\mu_d$ increases.

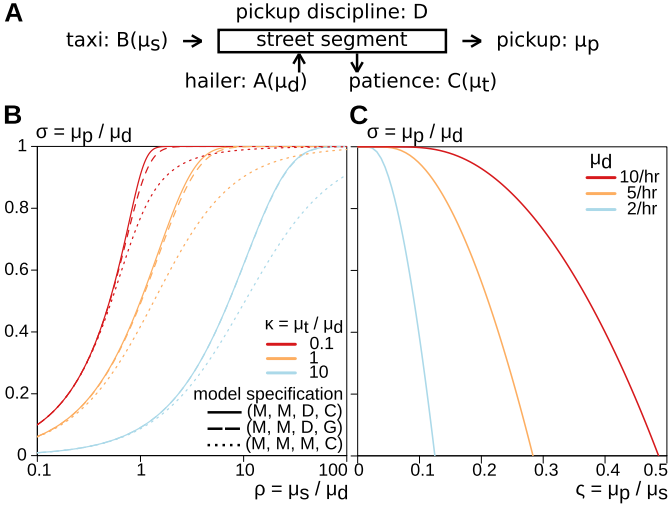

Fig. 1. Segment pickup model. (A) Model diagram. Hailers arrive at a oneway street segment as a stochastic process $A$ with rate $\mu_d$. Vacant taxis enter the segment as an independent stochastic process $B$ with rate $\mu_s$. Hailers quit waiting if not picked up within their patience, which is a random variable $C$ with expectation $1/\mu_t$. Pickup discipline $D$ determines which hailer gets picked up. (B) Dimensionless pickup rate function $\sigma(\rho, \kappa)$ under different model specifications and at constant cover numbers $\kappa$. (C) Pareto front of performance metrics (demand fulfillment $\sigma$ and supply realization $\varsigma$) with 4-minute service guarantee $T$, at different demand rates $\mu_d$.

For the rest of this subsection, we derive analytical forms of the pickup rate function for several model specifications.

In standard queueing theory, customers enter a system of several servers and request service; if every server is already serving a customer, the other customers line up in a queue and wait for free servers. In Kendall's notation [43], a particular type of queueing system is identified as $A/B/s$: customer inter-arrival times are independently distributed as $A$; service time for a customer is distributed as $B$; the number of servers is $s$. If not specified, queue capacity and population size both default to infinity. Common queueing disciplines include First In First Out (FIFO) and Service In Random Order (SIRO). Customer impatience has also been studied in queueing theory: balking refers to a customer voluntarily or involuntarily not joining the system, due to bounded system capacity [44] or at probabilities conditional on queue size or expected waiting time [45]; reneging refers to a customer leaving the system before starting or completing their service [46]. In the following, we use subscript $s$ for customer impatience on time within system (sojourn time), and use $w$ for customer impatience only in waiting-line. For example, $GI/G/s/FIFO + G_s$ refers to a queueing system with a general independent (GI) arrival process of customers, a general distribution (G) of service time, $s$ servers, vacant servers serving the first customer in queue, and a general distribution of customer impatience on sojourn time ($G_s$).

In general, our segment pickup model is not a standard queue because although the pickup discipline defines a queue of hailers, there is no dedicated server in the system. If the hailer at queue head quits before a vacant taxi arrives, the next hailer at queue head needs to wait for an extra amount of time that is distributed differently from the vacant taxi inter-arrival time $B$. But if $B$ is exponentially distributed (or memoryless), the time till pickup for the hailer at queue head will have the same exponential distribution, independent from the arrival times of passed vacant taxis and previous hailers. Thus, a $(GI, M, G, G)$ model coincides with a $GI/M/1/SIRO + G_s$ queue, referring to a general independent arrival process of hailers, Markovian/Poisson (M) arrival process of vacant taxis, a general distribution of hailer patience, and greedy pickup discipline. Similarly, a $(GI, M, G, C)$ model coincides with a $GI/M/1/FIFO + G_s$ queue, with courteous pickup discipline.

For models $(M, M, D, C)$ and $(M, M, D, G)$, where $D$ refers to a degenerate distribution (i.e. a deterministic number), both hailer and vacant taxi arrival processes are Poisson, and hailer patience is a constant. These models coincide with $M/M/1/FIFO + D_s$ and $M/M/1/SIRO + D_s$ queues respectively, which are single-server queues with Markovian inter-arrival time and service time, and with deterministic customer impatience on time in system. Reference [38], [39] first studied these two types of queues, and obtained closed expressions for the ratio of the average rate at which customers are lost to the average arrival rate. Because demand fulfillment is complement to "lost customer probability" [39] or "survival rate" [38], for model $(M, M, D, C)$, we have:

$$\sigma(\rho, \kappa) = \begin{cases} \rho \dfrac{e^{(\rho-1)/\kappa} - 1}{\rho e^{(\rho-1)/\kappa} - 1}, & (\rho \neq 1) \\ \dfrac{1}{\kappa + 1}, & (\rho = 1) \end{cases} \quad (1)$$

Similarly, for model $(M, M, D, G)$, we have:

$$\sigma = \rho - \rho\left(1 + \sum_{n=1}^{\infty} \rho^{-n} \prod_{m=1}^{n}\left(1 - e^{-\frac{\rho}{m\kappa}}\right)\right)^{-1} \quad (2)$$

For model $(M, M, M, C)$, both hailer and vacant taxi arrival processes are Poisson, and hailer patience is exponentially distributed. This model coincides with the $M/M/1/FIFO + M_s$ queue, which is studied as the Type II behavior of the unbounded queue in [40]. Because demand fulfillment is "the probability that an arrival completes service", for model $(M, M, M, C)$ we have:

$$\sigma = \rho - \rho \left[ 1 + \exp\left(\frac{1 + \rho \ln \kappa}{\kappa} + \ln \gamma \left(\frac{\rho}{\kappa} + 1, \frac{1}{\kappa}\right)\right) \right]^{-1} \quad (3)$$

Here $\gamma(a, x) = \int_0^x t^{a-1} e^{-t} \, dt$ is the lower incomplete gamma function.

Substituting the dimensionless numbers with their definitions gives the dimensional form of pickup rate functions. For model $(M, M, D, C)$, it is:

$$\mu_p(\mu_d, \mu_s, \mu_t) = \begin{cases} \mu_s \mu_d \dfrac{e^{\mu_s/\mu_t} - e^{\mu_d/\mu_t}}{\mu_s e^{\mu_s/\mu_t} - \mu_d e^{\mu_d/\mu_t}}, & (\mu_s \neq \mu_d) \\ \dfrac{\mu_s^2}{\mu_s + \mu_t}, & (\mu_s = \mu_d) \end{cases} \quad (4)$$

For model $(M, M, D, G)$:

$$\mu_p = \mu_s - \mu_s \left(1 + \sum_{n=1}^{\infty} \left(\frac{\mu_d}{\mu_s}\right)^n \prod_{m=1}^{n}\left(1 - e^{-\frac{\mu_s}{m\mu_t}}\right)\right)^{-1} \quad (5)$$

For model $(M, M, M, C)$:

$$\mu_p = \mu_s - \mu_s \left[1 + e^{\frac{\mu_d}{\mu_t}} \left(\frac{\mu_d}{\mu_t}\right)^{-\frac{\mu_s}{\mu_t}} \gamma\left(\frac{\mu_s}{\mu_t} + 1, \frac{\mu_d}{\mu_t}\right)\right]^{-1} \quad (6)$$

With a general probability model for hailer inter-arrival time and deterministic hailer patience, model $(GI, M, D, C)$ corresponds to queue $G/M/1 + D_s$ and also has an analytical form for the pickup rate function. Reference [41] solved the stationary waiting-time distribution function $W(x)$ for $M/G/1 + D_s$ queues. In his notation, $\{v_n\}$ is the difference between customer patience and actual queueing time, whose limiting distribution is $V(x)$. As he also noted, $\{v_n\}$ is the dual of waiting time $\{w_n\}$, so the functional form of $V(x)$ of a $G/M/1 + D_s$ queue is the same as $W(x)$ of an $M/G/1 + D_s$ queue. Thus, because demand fulfillment $\sigma = 1 - V(0)$, using definition $\sigma = \mu_p/\mu_d$ and expression of $W(x)$, for model $(GI, M, D, C)$ we have:

$$\mu_p = \mu_s - \mu_s \left( \sum_{n=0}^{\infty} \int_{0-}^{T} \frac{[-\mu_s(T-u)]^n}{n!} e^{\mu_s(T-u)} \, dA^{n*}(u) \right)^{-1} \quad (7)$$

Here $\mu_p$ is a function of $(\mu_s, A, T)$, where $A(x)$ is the distribution of hailer inter-arrival time and $T$ is hailer patience. $A^{n*}(x)$ is the $n$-fold convolution of $A(x)$ with itself, with $A^{0*}(x) = H(x)$, the Heaviside function.

### B. Equilibrium Supply Route Model

To estimate supply rate, we model the behavior of taxi drivers. Taxi drivers maximize their income. Since we assume they do not deny hailers, drivers cannot exploit hailers' destination to their advantage, so they will maximize their pickups per search time $\mu_{pi} = \sum_x \mu_{pix}$. Here subscript $i$ denotes driver, subscript $x$ denotes street segment, and we use a notational convention such that omitted subscripts indicate summation. To maximize $\mu_{pi}$, drivers allocate their search time $s_i$ over the street network, so the strategy vector is $\mathbf{s}_i = \{s_{ix}\}$. Because without other information, drivers do not know if any hailer is waiting on the segment they are about to search, we assume every pass of a vacant taxi on a given segment has equal probability of picking up a hailer, just like a Bernoulli trial. Thus, drivers' pickup rates on the segment are proportional to the supply rates they contribute. Formally, driver $i$ searching segment $x$ with supply rate $\mu_{six}$ has an expected pickup rate on this segment $\mu_{pix} = \mu_{six} \alpha_x$, where $\alpha_x$ is a proportionality constant of the segment. Summing up the equation we get $\sum_i \mu_{pix} = \sum_i \mu_{six} \alpha_x$, so that $\alpha_x = \mu_{px}/\mu_{sx}$, and thus $\mu_{pix} = \mu_{six} \mu_{px}/\mu_{sx}$. Given segment length $l_x$ and taxi search speed $\tilde{v}_x$, the search time driver $i$ spends on segment $x$ per unit time is $s_{ix} = \mu_{six} l_x / \tilde{v}_x$. So the expected pickup per search time on the segment is:

$$\frac{\mu_{pix}}{s_{ix}} = \frac{\mu_{six} \mu_{px}/\mu_{sx}}{\mu_{six} l_x / \tilde{v}_x} = \frac{\mu_{px} \tilde{v}_x}{\mu_{sx} l_x} \equiv w_x \quad (8)$$

As we can see, this value is not driver-specific, so we denote it as $w_x$. From our segment pickup model, we know that $\mu_{px}$ depends on $\mu_{sx}$, so $w_x$ is a function of $\mu_{sx}$, which is a function of search time $s_x$ on the segment. From Eq. 8, now we have $\mu_{pi} = \sum_x \mu_{pix} = \sum_x s_{ix} w_x(s_x)$. In other words, driver's objective $\mu_{pi}$ is a function of their strategy vector and the aggregate strategy vector: $\mu_{pi}(\mathbf{s}_i, \mathbf{s})$.

This effectively forms a game among taxi drivers, and the equilibrium of driver supply route choice can be formulated as follows: given the strategies of all taxi drivers, no driver can find a supply route with a higher pickup rate than the one already chosen. To estimate supply equilibrium, consider the drop-off locations of taxi rides as random, then at any moment vacant taxis starting to search for hailers are distributed on the street network according to drop-off frequency. These drivers will try to maximize their pickup rates, competing against each other and other vacant taxis still searching. These drivers do not know the time of their next pickup, so they will choose to search on segments nearby that have the highest expected pickup per search time $w_x$. If $w_x$ is not uniform, drivers close to a segment with high $w_x$ will move towards and search on it. But this increases supply rate $\mu_{sx}$ on the segment, and despite pickup rate $\mu_{px}$ on the segment would also increase as a result, it would be less than proportional due to the law of diminishing returns, so from Eq. 8, $w_x$ will decrease. When the influx of drivers to this segment reaches a level such that $w_x$ is no longer locally maximum, drivers will move on to search other segments to increase their pickup per search time $\mu_{pi}$. At equilibrium, the collective search strategy of taxi drivers will result in a network with uniform pickup per search time on segments being actively searched while other segments have lower pickup per search time, so that no driver can unilaterally change strategy for a faster expected pickup. This means the equilibrium supply distribution satisfies: $w_x = w \geq w_y, \forall x, y, s_x > 0$. Note that the constant $w$ is the expected pickup per search time at equilibrium, which can be estimated as the total pickup divided by total search time, $w = \mu_p/s$. From Eq. 8, our formula for estimating equilibrium supply distribution can now be expressed as:

$$\mu_{sx} = \frac{s \tilde{v}_x \mu_{px}}{l_x \mu_p} \quad (9)$$

### C. Summary and Solution Procedure

Now we provide a procedure for solving the supply and demand rates over a street network during a specific time window, given taxi trip records with vehicle identifier and the location and time of pickup and drop-off:

1) Pickup rates $\mu_{px}$: count the pickups on segemnt $x$, divide it by the duration $\Delta t$ of the time window.
2) Total search time $s$: sum up the time when active drivers are in between trips, divide it by $\Delta t$.
3) Taxi search speed $\tilde{v}_x$: estimated as the typical traffic speed, where the speed of each trip is computed as trip distance divided by trip duration.
4) Segment lengths $l_x$: lengths of the line strings of the street network.
5) Supply rates $\mu_{sx}$: by Eq. 9, using $\mu_{px}$, $s$, $\tilde{v}_x$, and $l_x$.
6) Demand rates $\mu_{dx}$: with $\mu_{px}$, $\mu_{sx}$, and $\mu_{tx}$, solve nonlinear equations $\mu_{px} = \mu_p(\mu_{dx}, \mu_{sx}, \mu_{tx})$ using root-finding algorithms such as Newton-Raphson or bisection, see Eq. 4–7. Hailer impatience $\mu_{tx}$ is a parameter.

Our procedure assumes Poisson arrivals of hailers and vacant taxis, equal pickup probability per search on a specific street segment, and that drivers maximize expected pickup per search time. The justification behind these assumptions have been explained earlier, and we validate the Poisson assumption in Subection IV-A. It might be better to assume that drivers maximize pickup rate over their search route, considering both drop-off and pickup locations; but the solution would be more complex and we leave it for future work. We note that, for computational simplicity, in our estimation we set taxi search speed $\tilde{v}_x$ to network average $\tilde{v}$. We justify this by considering that, different from taxis in trips, taxis searching for hailers would not drive at a pace that varies significantly over space. For papers that estimate segment-level traffic speed, see [21], [22]. We show in Subection IV-B that the overall procedure provides stable demand estimates.

## III. RESULTS

### A. Supply and Demand Distributions

To estimate taxi supply and demand distributions, we first identify time intervals that can be reasonably assumed as realizations of the same random field. Because transportation is essentially a social phenomenon, we partition each year into seasons based on social events. The observed annual patterns of taxi activity in NYC can be categorized into spring, summer, fall, and winter seasons, separated by federal holidays. Within each season we exclude certain days as exceptions: public holidays, custom, extreme weather, and days with significant data issues. Both spring and fall seasons have stable weekly pickup patterns, as most of the city's population are at work on weekdays. In this paper we choose the spring season for its regularity and duration, as we prefer a larger sample size for estimation. From Tuesday through Thursday, taxi pickup patterns within each day are almost identical. If we pick short time intervals of a day, variations in transportation activities become negligible. We may then regard each time interval as a cluster of observations of the same random field, independent from each other as they are distantly placed on the time axis. In this paper we choose the AM peak from 8am to 9am Tuesdays to Thursdays, as they have the most similar taxi and transportation activities. Given a time interval, we need to choose the relevant sub-sequences of trips for each taxi, to determine the actual pickups in this interval and the search efforts that lead to these pickups. This procedure is important because within a short time interval each taxi does not make many trips, careless counting may thus cause large error. Time sampling procedure is detailed in Subsection V-B.

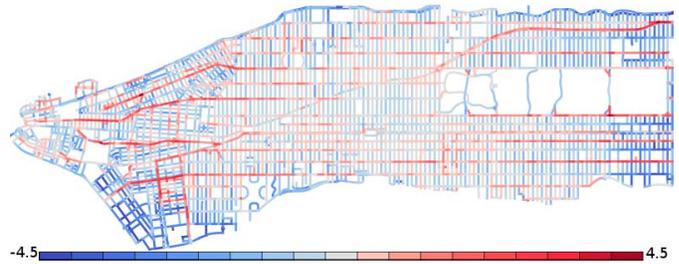

Fig. 2. Pickup-dropoff ratio of taxi trips in core Manhattan, log2 scale. Midtown has relatively balanced pickup and drop-off (grey); streets leading towards Midtown have much more pickups than drop-offs (red); many streets far from Midtown have much more drop-offs than pickups (blue).

With a subset of trip records, we match GPS locations onto street segments. To contain the size of the road network without truncating much of taxi activity, we choose a part of Manhattan where most taxi pickups are located, called "core Manhattan", defined as the Manhattan Island south of 130th Street. 92.5% of all taxi trips originate from core Manhattan, and 84.5% start and end within the region. We extract NYC road network from OpenStreetMap (OSM), and create a compressed graph using Open Source Routing Machine (OSRM) [47], where edges are street segments as we defined. The OSRM compressed graph of core Manhattan has 6,001 edges and 7,055 one-directional segments. GPS recordings are noisy but reasonably accurate, with degraded quality in densely built area due to urban canyon effects. Compared with the typical distance of 79 meters between street center lines in Manhattan, matching GPS locations to the nearest street segment would be correct in most cases except in downtown areas. About 2.2% trip records have missing pickup and drop-off GPS locations, and we consider the overall extent of GPS missing values as acceptable. Location matching is detailed in Subsection V-C. Taxi pickup distribution is spatially highly heterogeneous, while drop-off distribution is more spread out. The map of pickup-dropoff ratio (Fig. 2) suggests that taxi drivers prefer to head back to high demand area after a drop-off at a low demand location, consistent with our equilibrium supply model.

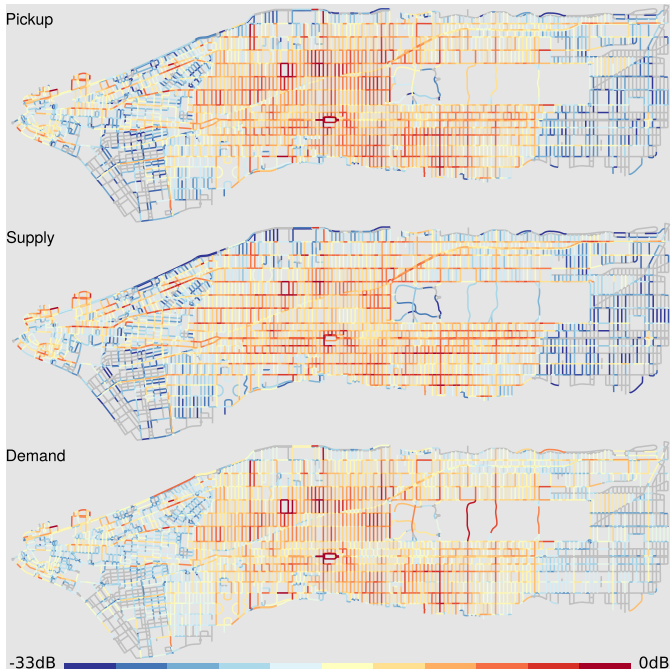

Fig. 3. Taxi pickup, supply, and demand distributions in core Manhattan, 8am-9am spring season 2012. Because taxi activities are highly heterogeneous over space, colors are in logarithmic scale (dB), relative to the highest values.

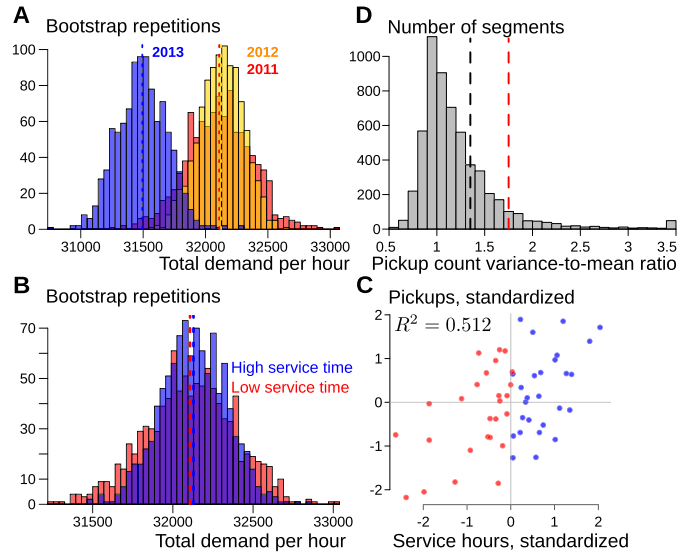

Fig. 4. ANOVA and Poisson tests, using trips in core Manhattan, spring, 8am–9am. Sampling distributions of total demand rate, 1000 bootstrap repetitions per group: (A) by year, 2011–2013; (B) by total taxi service hours, partitioning the 2012 sample into two equally sized groups shown in (C). (D) Histogram of variance-to-mean ratio (VMR) of pickup counts on street segments in 2012 (sample size N = 53). Pickup counts on two-way segments are considered as uncorrelated and tested as one unit. Values larger than 3.5 are clipped to 3.5. Dashed lines show thresholds of the sampling distribution of Poisson VMR, with p-values 0.05 (black) and 0.001 (red).

We apply our models to estimate the supply and demand distributions in core Manhattan. Based on taxi trip distance and duration, the typical traffic speed in core Manhattan during the 8am-9am peak hour is about 14.5 km/h (9 mph), which we use in supply estimation. We assume taxi search speed is constant as drivers focus on curbside hailers and do not compete with traffic. We estimate demand with our preferred pickup model: $(M, M, M, C)$ with 4-minute hailer mean patience (see Subsection IV-C for justification). Results for spring season 2012 are shown in Fig. 3, along with pickup distribution for comparison. Taxi pickup overall concentrates around Midtown, with spatial variations; taxi supply instead is mostly along the numbered avenues, yet not uniform; taxi demand is more spread out than pickup, but the hot spots are similar to those of pickups. Demand estimates for each hour of a weekday in each season in 2012 is provided as an animation in Attachments.

Street-hail taxi performs very well in core Manhattan, where about half of the segments have at least five hailers per hour. At demand rate 5/hr, Fig. 1C shows that pickup can be guaranteed within 4 minutes for 95% of the hailers, while one in ten vacant taxi passes are successful. Compare with TNCs, where average wait time is estimated to be 3-4 minutes in 2017 [2], taxis perform better, especially in Midtown where demand is the highest.

### B. Decline of Taxi Demand

We compare taxi demand in the years of 2011-2013, see Fig. 4A. While taxi demand in 2011 and 2012 are almost the same, it declined about 2% in 2013. Table I summarizes service time and pickup, supply, and demand rates in core Manhattan in spring season 2012; results for 2011 and 2013 are shown as percentage change relative to 2012. In the table, $R^2$ refers to the coefficient of determination between service time and pickups. CV stands for the coefficient of variation of the corresponding estimate, computed using bootstrap standard error, also shown in percentages. As with the results on high and low service times in 2012 (Fig. 4B), variation in demand estimates is higher if supply level is lower. It is clear from the table that the 0.6% fewer trips made in 2011 can be completely accounted for by the 3.8% less supply rates; while the 2% reduction in pickups in 2013 was solely due to the 2% decline in demand rates. Effects in both years are statistically significant.

TABLE I
TAXI ACTIVITY IN CORE MANHATTAN, 8AM-9AM SPRING SEASONS

| Year | 2012 | 2011 | 2013 |
|---|---|---|---|
| Sample size | 53 | 43 | 44 |
| Service time | 7708.31 hrs/hr | -2.32% | 0.394% |
| Pickup rate | 26920.4 pickups/hr | -0.628% | -1.97% |
| ($R^2$) | (0.512) | (0.302) | (0.477) |
| Supply rate | 298924 passes/hr | -3.79% | -0.494% |
| (CV) | (1.29%) | (1.70%) | (1.57%) |
| Demand rate | 32124.5 hailers/hr | -0.0461% | -1.965% |
| (CV) | (0.546%) | (0.808%) | (0.636%) |

With a fixed taxi supply, a stable economy, and no major change in transportation infrastructure, it should be expected that taxi demand in NYC should have stayed the same through those years. The decline in taxi demand in 2013 may be caused by two factors: the entry of Uber in NYC, and the Taxi and Limousine Commission (TLC) fare raise, both in the second half of 2012. In particular, Uber announced UberX on 2012-07-04, a service using hybrid vehicles and more affordable than Uber Black, its black car service. Separately, TLC passed rules effective on 2012-09-04 which increased metered fare by 25% and also increased the flat fare and surcharge of airport trips, on average raising trip fare by 17%. However, the actual cause of decline is beyond the scope of the current paper. From another perspective, the result that taxi demand is the same in 2011 and 2012 reaffirmed our assumption that urban transportation is in dynamic equilibrium in the absence of systematic changes.

We note that, the TLC e-hail (mobile app) pilot program did not affect the street-hailing nature of taxis. TLC started an e-hail pilot program on 2013-04-26, which was interrupted from 2013-05-01 to 2013-06-06 due to litigation from an appellate judge. In late 2013 average daily e-hail requests is under 5,000 and fulfillment rate is about 30%, resulting in only 0.3% taxi trips. In October 2016, only 28,281 taxi trips were originated via e-hail, less than monthly average in late 2013. Taxi trips requested from the mobile app are also in the trip records.

## IV. Discussion

### A. Poisson Assumption

Here we test if the arrivals of hailers and vacant taxis can be assumed to be Poisson processes. Although the arrivals of hailers and vacant taxis are not directly observed, if they are in fact two independent Poisson processes, our Monte Carlo simulation shows that the resulting pickups are close to another Poisson process. Thus we continue to test if the observed pickup counts are Poisson.

Overdispersion is a common issue in count data where the variation is larger than a standard Poisson model would suggest, which arises when the arrivals are in clusters. This issue has long been discussed in statistical literature, with many tests proposed for it, see for example [48]. For NYC taxi trip records, [18] suggests that taxi pickups are highly overdispersed, with variances on the order of 10000 times larger than the averages. We calculate the variance-to-mean ratio (VMR) of pickup counts on street segments in core Manhattan, 8am-9am spring season 2012, shown in Fig. 4D. With variance-mean relationship specified as $\sigma^2 = \alpha\mu$, the null hypothesis $H_0 : \alpha = 1$ and the alternative hypothesis $H_1 : \alpha > 1$, the null is rejected at size 5% if the VMR exceeds 1.34455, and at size 0.1% if its exceeds 1.75. For the street segments in core Manhattan, the median VMR is 1.10, with 24.8% of the segments with VMR larger than 1.34455, and 8.06% of the segments with VMR larger than 1.75. Thus the Poisson assumption is consistent with observations on most of the segments, and would still be appropriate on most of the remaining segments. Street segments with very high pickup rates have higher VMR, which can be caused by occasional events that draw large crowds increasing taxi activity. On segments with taxi stands, typically at major transportation hubs, vacant taxis line up and wait for customers, which challenges our model assumption. These segments can be seen to have infinite supply rates as long as the taxi line is not empty, and thus demands are always fulfilled. Our model naturally handles this situation as the estimated equilibrium supply rate is proportional to pickup rate, and high supply rate is associated with high demand fulfillment, providing appropriate demand estimates even if the model assumptions are challenged. Our result on overdispersion stands in contrast to other literature suggested, which highlights the importance of time sampling and spatial unit selection.

### B. Stability of Demand Estimates

Since the estimated taxi demand distribution has never been directly measured, we validate it by testing its stability at different supply levels. Total pickup is positively correlated with taxi service hours in the same time interval, see Fig. 4C. The $R^2$ is close to 0.5, because both supply and demand levels contribute to the variance of pickup counts; $R^2$ has similar values in other years, see Table I. Here we use service hours to measure supply level, rather than taxi counts or search time, because not all taxis provide the same amount of service time that is in core Manhattan, while search time is confounded by taxi counts. Specifically, two competing factors affect the correlation between search time and pickups: assuming the average trip duration is stable, a fixed number of active taxis means a fixed sum of search and trip time, search time is thus negatively correlated with pickups; with more active taxis and a fixed demand, pickups and search time both increase, thus positively correlated. Assuming demand distribution is the same in all observations, the demand distribution estimate is stable if it is uncorrelated with supply level. In other words, demand estimates do not change by clustering observations of similar supply levels. We partition the original sample into two equally sized subsamples by service hours, and estimate the demand distribution separately with 1000 bootstrap resamples, results shown in Fig. 4B. The average demand estimates are very close, and the difference is not statistically significant. It means that the difference in pickups between the two subsamples are completely explained by the difference in supply levels, and the demand estimate is the same, consistent with our assumption.

We formalize this test of demand estimate stability and compare our model with other matching functions. Results are shown in Table II. In the table we use a nonparametric Behrens-Fisher t-test [49] for the mean values of total demand estimates using subsamples of high and low supply levels. None of the previous matching functions provide stable demand estimates— as their p-values are far less than 1, while our model shows no statistically significant difference. We note that while the stability of demand estimate is a desirable property, it is not the only justification of our model over previous methods, see other columns listed in the table and Section I.

### C. Choice of Model Specification

Here we discuss why we pick our preferred pickup model to be $(M, M, M, C)$ with 4-minute hailer mean patience on all street segments.

Recall that our segment-level pickup model admits a specification of the form $(A, B, C, D)$, and we need to specify that for each segment in the road network. We choose among the specifications $(M, M, M, C)$, $(M, M, D, C)$, $(M, M, D, G)$, and $(GI, M, D, C)$, which we have obtained analytical forms of their pickup rate functions, see Eq. 4–7. As discussed in Subsection IV-A, the Poisson assumption is consistent with observations on most of the segments, and is appropriate on most of the remaining segments, so we let $(A, B) = (M, M)$. This allows us to drop $(GI, M, D, C)$ for the more specific $(M, M, D, C)$. Comparing $(M, M, D, C)$ and $(M, M, D, G)$, the effect of greedy and courteous pickup disciplines appear to have little effect on pickup rate, see Fig. 1B, so we may regard them as the same. This leaves us with $(M, M, M, C)$ and $(M, M, D, C)$, i.e. either deterministic or exponential distribution of hailer patience $T$ on a segment.

TABLE II
COMPARISON OF TAXI–PASSENGER MATCHING FUNCTIONS

| Reference | Matching function | Type | Spatial unit | Origin | T statistic | p-value |
|---|---|---|---|---|---|---|
| [29] | $\mu_p = \min\{\mu_s, \phi\mu_d\}$ | aggregate | point | point matching | 194.026 | 0 |
| [12] | $\mu_p = AN_s^{\alpha_1} N_d^{\alpha_2}$ | aggregate | point, area | Cobb-Douglas production | -29.060 | 0 |
| [34] | $\mu_p = \mu_s\left[1 - \exp(-\alpha\mu_d/\mu_s)\right]$ | aggregate | area | urn-ball matching | 80.885 | 0 |
| This paper | Eq. 6 for model $(M, M, M, C)$ | mechanistic | segment | meeting on segments | 1.632 | 0.103 |

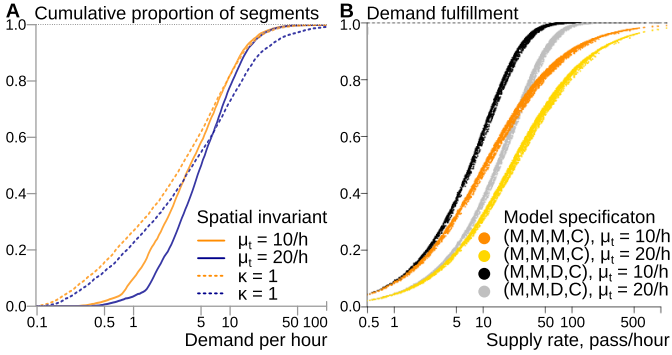

Fig. 5. Comparison of model specifications. (A) Cumulative distribution of demand rates over the street network of core Manhattan, estimated using $(M, M, M, C)$. The result is similar using $(M, M, D, C)$, and is left out for clarity. (B) Demand fulfillment vs. supply rate on street segments in core Manhattan, assuming constant impatience on all segments. As expected, demand fulfillment decreases if impatience is higher (lighter shades) or if hailer patience is exponentially distributed (orange) rather than deterministic (black).

To help determine hailer mean patience on each segment, we regard either impatience $\mu_t$ or cover number $\kappa = \mu_t/\mu_d$ as spatially invariant, i.e. constant across all segments of the network. Fig. 5A quantifies the effect of spatial invariant on demand estimate. For homogeneous impatience, we pick 10 and 20 per hour, which means an average hailer would wait for 6 or 3 minutes before quit waiting. For homogeneous cover number, we pick 1 and 3, which means hailer mean patience is equal to one or one third of the expected arrival time of the next hailer. The cases where cover number is spatially invariant have much more segments with very small and very large demand. Assuming cover number to be spatially invariant is inappropriate, as it would either make hailers on high demand segments too impatient or those on low demand segments too patient, oftentimes both. With cover number 1, hailers on a segment with demand rate 60 per hour have 1 minute patience on average, while those on segments with demand rate 1 per hour would have 60 minutes of patience on average. This would cause underestimates on low demand segments, and overestimates on high demand segments, which explains our result. We instead choose impatience to be the spatial invariant.

Specifications $(M, M, M, C)$ and $(M, M, D, C)$ are compared in Fig. 5B, with impatience $\mu_t$ as the spatial invariant. We see that demand fulfillment $\sigma$ roughly follows an S-curve of supply rate $\mu_s$, despite that it also depends on supply-pickup ratio $\rho$ which appears to have little influence. We prefer $(M, M, M, C)$, which means exponential distribution of hailer patience, because it allows individuals to differ in travel decisions. In Section III, we use the intermediate value of $\mu_t$, 15 per hour, which means hailer mean patience is 4 minutes.

## V. MATERIALS AND METHODS

### A. NYC Taxi Trip Records

The Taxicab Passenger Enhancement Program (TPEP) enables electronic collection and submission of NYC taxi trip records. TLC releases TPEP records to the public persuant to the Freedom of Information Law of New York State. We process the TPEP records from 2009 to 2013, the first five calendar years since TPEP systems were installed in all 13,237 Medallion taxis. These records constitute all the taxi trips in NYC during the same period until the launch of green cabs on 2013-08-08. TPEP trip records include time stamps and GPS positions of each taxi pickup and drop-off, along with other attributes. The original and processed data are available for reuse at [50].

### B. Time Sampling

Although we have five years of trip records, no all the years have acceptable data qualities. One of the TPEP vendors discontinued contract with TLC in 2010, resulting in incomplete data reporting. The 2009 data also has its own issues as the early stage of the TPEP program. Thus we focus our analysis on the 2011-2013 data. Based on the observed patterns, we categorize annual taxi activity into four seasons: spring season from Martin Luther King Jr. (MLK) Day to Memorial Day; summer season to Labor Day; fall season to Thanksgiving; and winter season to MLK Day next year. For an hour-long interval like 8am-9am, a trip is counted as within the time interval if the pickup time stamp is on or after 8am and before 9am. To count supply time, we sort the trip sequences of each taxi by pickup time stamps, and convert the trip records into search records by linking the drop-off attributes with those of the next pickup. We then subset the search records to those overlapping with the chosen time interval, and clip them to the bounds if any extends beyond the interval.

*C. Road Network and Map Matching*

For the road network of NYC, we filter OpenStreetMap (OSM) data for the public non-freeway vehicular road network. Specifically, we include OSM ways whose highway tag take one of the following values: trunk, primary, secondary, tertiary, unclassified, residential. We excluded motorway because no taxi pickup nor drop-off shall be on motorways. To make the road network strongly connected, we removed tunnels, bridges, and link roads. Rarely used values are excluded, such as road and living_street; so are roads not accessible by taxis, such as footway and service. The filtered OSM map has 8928 locations and 11458 edges. We use Open Source Routing Machine (OSRM) to create a compressed graph of 6001 edges. We exploit another module in OSRM to match GPS locations to the nearest segment, where longitudes and latitudes are transformed in Mercator projection for isotropic local scales of distance. Locations matched on two-way streets are assigned equally to both one-directional segments. The modified code is available at https://github.com/rudazhan/osrm-backend.

ACKNOWLEDGMENT

The authors would like to thank Abhishek Nagaraj of UC Berkeley and Henry S. Farber of Princeton University for sharing NYC taxi trip data, OpenStreetMap contributors for NYC map data, and Open Source Routing Machine contributors for graph preparation and map matching modules.

REFERENCES

[1] Mayor's Office of Operations, "Mayor's management report: Fiscal 2017," Mayor's Office Oper., New York, NY, USA, Tech. Rep. mmr2017, Sep. 2017.
[2] B. Schaller, "Empty seats, full streets: Fixing manhattan's traffic problem," Schaller Consulting, Brooklyn, NY, USA, Tech. Rep., Dec. 2017. [Online]. Available: http://www.schallerconsult.com/rideservices/emptyseats.pdf
[3] P. Santi, G. Resta, M. Szell, S. Sobolevsky, S. H. Strogatz, and C. Ratti, "Quantifying the benefits of vehicle pooling with shareability networks," *Proc. Nat. Acad. Sci. USA*, vol. 111, no. 37, pp. 13290–13294, 2014.
[4] K. I. Wong, S. C. Wong, and H. Yang, "Modeling urban taxi services in congested road networks with elastic demand," *Transp. Res. B, Methodol.*, vol. 35, no. 9, pp. 819–842, 2001.
[5] H. Yang, S. C. Wong, and K. Wong, "Demand–supply equilibrium of taxi services in a network under competition and regulation," *Transp. Res. B, Methodol.*, vol. 36, no. 9, pp. 799–819, Nov. 2002.
[6] K. Zheng, Y. Zheng, X. Xie, and X. Zhou, "Reducing uncertainty of low-sampling-rate trajectories," in *Proc. ICDE*, Apr. 2012, pp. 1144–1155.
[7] R. Silva, S. M. Kang, and E. M. Airoldi, "Predicting traffic volumes and estimating the effects of shocks in massive transportation systems," *Proc. Nat. Acad. Sci. USA*, vol. 112, no. 18, pp. 5643–5648, 2015.
[8] P. Deville, C. Song, N. Eagle, V. D. Blondel, A.-L. Barabási, and D. Wang, "Scaling identity connects human mobility and social interactions," *Proc. Nat. Acad. Sci. USA*, vol. 113, no. 26, pp. 7047–7052, 2016.
[9] S. Jiang, Y. Yang, S. Gupta, D. Veneziano, S. Athavale, and M. C. Gonzalez, "The TimeGeo modeling framework for urban mobility without travel surveys," *Proc. Nat. Acad. Sci. USA*, vol. 113, no. 37, pp. E5370–E5378, 2016.
[10] X. Qian and S. V. Ukkusuri, "Taxi market equilibrium with third-party hailing service," *Transp. Res. B, Methodol.*, vol. 100, pp. 43–63, Jun. 2017.
[11] H. Yang and S. C. Wong, "A network model of urban taxi services," *Transp. Res. B, Methodol.*, vol. 32, no. 4, pp. 235–246, May 1998.
[12] H. Yang, C. W. Y. Leung, S. Wong, and M. G. H. Bell, "Equilibria of bilateral taxi–customer searching and meeting on networks," *Transp. Res. B, Methodol.*, vol. 44, nos. 8–9, pp. 1067–1083, 2010.
[13] S. Phithakkitnukoon, M. Veloso, C. Bento, A. Biderman, and C. Ratti, "Taxi-aware map: Identifying and predicting vacant taxis in the city," in *Proc. Int. Joint Conf. Ambient Intell.*, 2010, pp. 86–95.
[14] N. J. Yuan, Y. Zheng, L. Zhang, and X. Xie, "T-finder: A recommender system for finding passengers and vacant taxis," *IEEE Trans. Knowl. Data Eng.*, vol. 25, no. 10, pp. 2390–2403, Oct. 2013.
[15] H.-W. Chang, Y.-C. Tai, and Y.-J. Hsu, "Context-aware taxi demand hotspots prediction," *Int. J. Bus. Intell. Data Mining*, vol. 5, no. 1, pp. 3–18, 2010.
[16] L. Moreira-Matias, J. Gama, M. Ferreira, J. Mendes-Moreira, and L. Damas, "Predicting taxi-passenger demand using streaming data," *IEEE Trans. Intell. Transp. Syst.*, vol. 14, no. 3, pp. 1393–1402, Sep. 2013.
[17] W. Tu, Q. Li, Z. Fang, S. Shaw, B. Zhou, and X. Chang, "Optimizing the locations of electric taxi charging stations: A spatial–temporal demand coverage approach," *Transp. Res. C, Emerg. Technol.*, vol. 65, pp. 172–189, Apr. 2016.
[18] C. Yang and E. J. Gonzales, "Modeling taxi demand and supply in New York city using large-scale taxi GPS data," in *Seeing Cities Through Big Data*, P. Thakuriah, N. Tilahun, and M. Zellner, Eds. Cham, Switzerland: Springer, 2017, pp. 405–425.
[19] D. Shao, W. Wu, S. Xiang, and Y. Lu, "Estimating taxi demand-supply level using taxi trajectory data stream," in *Proc. IEEE ICDMW*, Nov. 2015, pp. 407–413.
[20] P. S. Castro, D. Zhang, and S. Li, "Urban traffic modelling and prediction using large scale taxi GPS traces," in *Pervasive Computing*. Berlin, Germany: Springer, 2012, pp. 57–72.
[21] X. Zhan, S. Hasan, S. V. Ukkusuri, and C. Kamga, "Urban link travel time estimation using large-scale taxi data with partial information," *Transp. Res. C, Emerg. Technol.*, vol. 33, pp. 37–49, Aug. 2013.
[22] X. Wang *et al.*, "Speed variation during peak and off-peak hours on urban arterials in Shanghai," *Transp. Res. C, Emerg. Technol.*, vol. 67, pp. 84–94, Jun. 2016.
[23] J. Alonsomora, S. Samaranayake, A. Wallar, E. Frazzoli, and D. Rus, "On-demand high-capacity ride-sharing via dynamic trip-vehicle assignment," *Proc. Nat. Acad. Sci. USA*, vol. 114, no. 3, pp. 462–467, 2017.
[24] Y. Ge, H. Xiong, A. Tuzhilin, K. Xiao, M. Gruteser, and M. Pazzani, "An energy-efficient mobile recommender system," in *ACM KDD*, Jul. 2010, pp. 899–908.
[25] J. W. Powell, Y. Huang, F. Bastani, and M. Ji, "Towards reducing taxicab cruising time using spatio-temporal profitability maps," in *Advances in Spatial and Temporal Databases*. Berlin, Germany: Springer, 2011, pp. 242–260.
[26] M. Qu, H. Zhu, J. Liu, G. Liu, and H. Xiong, "A cost-effective recommender system for taxi drivers," in *Proc. ACM KDD*, Aug. 2014, pp. 45–54.
[27] S. Qian, J. Cao, F. L. Mouël, I. Sahel, and M. Li, "SCRAM: A sharing considered route assignment mechanism for fair taxi route recommendations," in *Proc. ACM KDD*, Aug. 2015, pp. 955–964.
[28] X. Zhan, X. Qian, and S. V. Ukkusuri, "A graph-based approach to measuring the efficiency of an urban taxi service system," *IEEE Trans. Intell. Transp. Syst.*, vol. 17, no. 9, pp. 2479–2489, Sep. 2016.
[29] R. Lagos, "An alternative approach to search frictions," *J. Political Economy*, vol. 108, no. 5, pp. 851–873, 2000.
[30] H. Yang and T. Yang, "Equilibrium properties of taxi markets with search frictions," *Transp. Res. B, Methodol.*, vol. 45, no. 4, pp. 696–713, 2011.
[31] X. Qian and S. V. Ukkusuri, "Time-of-day pricing in taxi markets," *IEEE Trans. Intell. Transp. Syst.*, vol. 18, no. 6, pp. 1610–1622, Jun. 2017.
[32] M. Ramezani and M. Nourinejad, "Dynamic modeling and control of taxi services in large-scale urban networks: A macroscopic approach," *Transp. Res. C, Emerg. Technol.*, vol. 23, pp. 41–60, 2017.
[33] G. R. Fréchette, A. Lizzeri, and T. Salz, "Frictions in a competitive, regulated market: Evidence from taxis," *Amer. Econ. Rev.*, vol. 109, pp. 2954–2992, Aug. 2016.
[34] N. Buchholz, "Spatial equilibrium, search frictions and dynamic efficiency in the taxi industry," Mimeo, Princeton Univ., Princeton, NJ, USA, 2019. [Online]. Available: https://scholar.princeton.edu/sites/default/files/nbuchholz/files/buchholz_taxi_2018.pdf
[35] K. Burdett, S. Shi, and R. Wright, "Pricing and matching with frictions," *J. Political Economy*, vol. 109, no. 5, pp. 1060–1085, 2001.
[36] G. R. Butters, "Equilibrium distributions of sales and advertising prices," *Uncertainty Econ.*, vol. 44, pp. 495–513, 1978.
[37] D. G. Kendall, "Some problems in the theory of queues," *J. Roy. Statist. Soc. B, Methodol.*, vol. 13, no. 2, pp. 151–185, 1951.

[38] D. Y. Barrer, "Queuing with impatient customers and indifferent clerks," *Oper. Res.*, vol. 5, no. 5, pp. 644–649, 1957.
[39] D. Y. Barrer, "Queuing with impatient customers and ordered service," *Oper. Res.*, vol. 5, no. 5, pp. 650–656, 1957.
[40] C. J. Ancker and A. Gafarian, "Queueing with impatient customers who leave at random," *J. Ind. Eng.*, vol. 13, nos. 84–90, pp. 171–172, Mar. 1962.
[41] D. J. Daley, "Single-server queueing systems with uniformly limited queueing time," *J. Austral. Math. Soc.*, vol. 4, no. 4, p. 489, Nov. 1964.
[42] S. Zhang and Z. Wang, "Inferring passenger denial behavior of taxi drivers from large-scale taxi traces," *PLoS ONE*, vol. 11, no. 11, Nov. 2016, Art. no. e0165597.
[43] D. G. Kendall, "Stochastic processes occurring in the theory of queues and their analysis by the method of the imbedded Markov chain," *Ann. Math. Statist.*, vol. 24, no. 3, pp. 338–354, 1953.
[44] F. A. Haight, "Queueing with balking," *Biometrika*, vol. 44, nos. 3–4, pp. 360–369, Dec. 1957.
[45] F. A. Haight, "Queueing with balking. II," *Biometrika*, vol. 47, nos. 3–4, pp. 285–296, Dec. 1960.
[46] F. A. Haight, "Queueing with reneging," *Metrika*, vol. 2, no. 1, pp. 186–197, Dec. 1959.
[47] D. Luxen and C. Vetter, "Real-time routing with openstreetmap data," in *Proc. ACM SIGSPATIAL GIS*, Nov. 2011, pp. 513–516.
[48] A. C. Cameron and P. K. Trivedi, "Regression-based tests for overdispersion in the Poisson model," *J. Econometrics*, vol. 46, no. 3, pp. 347–364, 1990.
[49] F. Konietschke, M. Placzek, F. Schaarschmidt, and L. A. Hothorn, "nparcomp: An R software package for nonparametric multiple comparisons and simultaneous confidence intervals," *J. Stat. Softw.*, vol. 64, no. 9, pp. 1–17, 2015.
[50] R. Zhang. (Feb. 2018). *New York City Taxi Trip Records, (2009–2013)*. [Online]. Available: https://osf.io/zhp7k

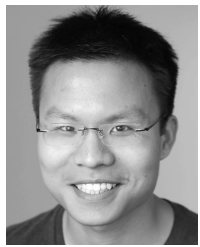

**Ruda Zhang** received the B.E. degree from Peking University, Beijing, China, in 2012, and the M.A. degree in economics and the Ph.D. degree in civil engineering from the University of Southern California, Los Angeles, CA, USA, in 2018. He is currently a Post-Doctoral Scholar with the University of Southern California. His research interests include urban systems, transportation, sensor data analytics, game theory, and institutional analysis.

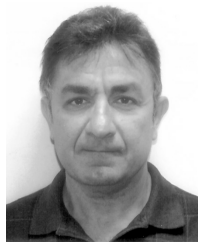

**Roger Ghanem** received the B.E. degree from the American University of Beirut, in 1984, and the master's and Ph.D. degrees from Rice University, in 1985 and 1989, respectively. He is currently the Gordon S. Marshall Professor of engineering technology with the Department of Civil and Environmental Engineering, University of Southern California. His research is in the area of computational stochastic mechanics and uncertainty quantification with focus on coupled, heterogeneous, and multiscale systems. He is a fellow of USACM, WCCM, EMI, and AAAS. He currently serves on the Executive Council of the U.S. Association for Computational Mechanics (USACM) and as a Chair of the Uncertainty Quantification SIAG of SIAM/SAS.